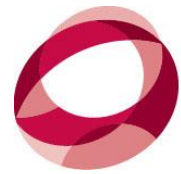

# Computing Research for the Climate Crisis

**Authors:** Nadya Bliss (Arizona State University), Elizabeth Bradley (University of Colorado Boulder), Claire Monteleoni (University of Colorado Boulder)

**Additional contributions by**: Ilkay Altintas (San Diego Supercomputer Center), Kyri Baker (University of Colorado Boulder), Sujata Banerjee (VMware), Andrew A. Chien (University of Chicago), Thomas Dietterich (Oregon State University), Ian Foster (Argonne National Labs), Carla P. Gomes (Cornell University), Chandra Krintz (University of California, Santa Barbara), Jessica Seddon (World Resources Institute), and Regan Zane (Utah State University)

## Introduction

Climate change is an existential threat to the United States and the world. Inevitably, computing will play a key role in mitigation, adaptation, and resilience in response to this threat. The needs span all areas of computing, from devices and architectures (e.g., low-power sensor systems for wildfire monitoring) to algorithms (e.g., predicting impacts and evaluating mitigation), and robotics (e.g., autonomous UAVs for monitoring and actuation)—as well as every level of the software stack, from data management systems and energy-aware operating systems to hardware/software co-design. There are a number of cross-cutting threads here. Artificial intelligence (AI), for instance, can be used to identify novel materials, plan trajectories for agricultural monitoring robots, and increase awareness of climate change among the general public. The growing computational needs of the modeling, simulation, and machine-learning technologies that will be required to support our response to climate change require special attention, as they themselves can contribute to the problem. Sociotechnical computing also cuts across the problems posed by climate change, playing an important role in representative datasets, trustworthy computing, equitable innovation, critical inquiry, and multidisciplinarity; see also [1].

*The goal of this white paper is to highlight the role of computing research in addressing climate change-induced challenges.* To that end, we outline *six key impact areas* in which these challenges will arise—energy, environmental justice, transportation, infrastructure, agriculture, and environmental monitoring & forecasting—then identify specific ways in which computing research can help address the associated problems. These impact areas will create a driving force behind, and enable, cross-cutting, system-level innovation. We further break down this information into *four broad areas of computing research*: devices & architectures, software, algorithms/AI/robotics, and sociotechnical computing. Some examples of specific instantiations of technologies from these four areas in each of the six impact areas appear below:

---

[1] E. Mynatt et al. *Harnessing the Computational and Social Sciences to Solve Critical Societal Problems.* National Science Foundation*, 2021.* https://www.nsf.gov/cise/advisory/SBE-CISE-roundtable-report.pdf

Energy
- *Hardware*: Sensors and sensor networks for monitoring power generation and consumption; energy-harvesting devices
- *Software*: Managing, cleaning, fusing, and distributing heterogeneous data from multiple sources
- *AI/Robotics/Algorithms*: Planning, optimization, and decision support for production, distribution, and consumption of energy; AI-enabled materials science for renewables
- *Sociotechnical*: Appropriate cost functions for optimization; communicating about decisions

Environmental justice
- *Software*: Rich data sets and models that properly capture and expose equity aspects, hidden biases, and economic factors
- *AI/Robotics/Algorithms*: Modeling and decision-support strategies that leverage those data and manage cascading risks
- *Sociotechnical*: Critical inquiry methodology, question- (vs. solution-) driven science, and stakeholder engagement

Transportation
- *Hardware*: Sensors and sensor networks for monitoring salient variables (traffic, goods, people, pollution, energy availability & needs, etc.)
- *Software*: Managing and disseminating data; creating effective models that couple those variables with economic forces and constraints
- *AI/Robotics/Algorithms*: Spatiotemporal planning strategies to optimize the routing of flows in the network
- *Sociotechnical*: Helping people understand how and why to act in this new system

Infrastructure
- *Hardware*: Sensors and sensor networks for monitoring roads, bridges, communication networks, etc.
- *Software*: Smart database management for lifetime of materials
- *AI/Robotics/Algorithms*: Optimization and decision support of flows of energy, goods, water, vehicles, people, power, etc.; AI-enabled materials science for green materials
- *Sociotechnical*: Equitable distribution and access to new technology (e.g. renewable energy, grid resiliency, power); stakeholder engagement

Agriculture
- *Hardware*: Sensors and sensor networks for monitoring water, temperature, crop growth, etc.
- *Software*: Systems for deployment and control of autonomous vehicles (e.g., UAVs)
- *AI/Robotics/Algorithms*: Algorithms that leverage rich sensor data, together with real-time information about economic factors and transportation networks, for planning and risk assessment
- *Sociotechnical*: Stakeholder engagement

Environmental monitoring and forecasting
- *Hardware*: Sensors and sensor networks for monitoring temperature, CO2 levels, ice coverage, etc.
- *Software*: Detection and diagnosis of sensor failure; managing assimilation of data into models
- *AI/Robotics*: Uncertainty quantification; system-level, risk-sensitive modeling, planning, and optimization strategies for climate variables, at all scales
- *Sociotechnical*: Ensuring saliency, credibility, and legitimacy in decision support systems; see also [2]

This list is not exhaustive; our aim here is not to provide a comprehensive treatment of either impact areas or computing research opportunities, but rather to highlight the vital role of computing research in climate change response. In the following six sections, we describe some of the problems that will arise in each of the impact areas in a bit more depth and motivate specific computing research challenges that will be involved in solving those problems. To succeed, this work cannot proceed in separate research silos, as it will demand advances both in individual research areas and in systems-level approaches that combine solutions from all four of those areas.

**Energy**

Energy system failures do not only have economic impact, but also human costs—up to and including loss of life. These harms are poised to worsen with the increased frequency and magnitude of extreme events due to climate change (hurricanes, heat waves, large rainfall events, etc.). Outright failures are not the only issue; normal operating conditions for these systems are changing rapidly as renewables and other "edge" sources are added to the generation mix and devices like electric cars become more widespread. Increasingly variable supply and demand—and the incorporation of a full accounting of the true costs of energy generation and supply—will require a complete rethinking of energy systems modeling. Planning, optimization, and control strategies are only as good as the models they use, and current models are not up to the coming task. In the face of the challenges posed by climate change, we need systems-level spatiotemporal models that combine traditional power-systems models (e.g., of generation, transmission, and distribution) with predictive demand models that incorporate economics, weather, and representations of hidden and downstream costs, as well as the supply chain issues that go into the problem. Critically, these models will need to be adaptive, which will require real-time monitoring and data assimilation, including fusion of data of multiple types and scales. Powerful systems-level models like this will enable effective optimization and decision-support strategies for efficient, adaptable, resilient energy systems:

---

[2] David Cash, William C. Clark, Frank Alcock, Nancy M. Dickson, Noelle Eckley, Jill Jäger. Salience, Credibility, Legitimacy and Boundaries: Linking Research, Assessment and Decision Making. KSG Working Papers Series, 2003.
https://dash.harvard.edu/bitstream/handle/1/32067415/Salience_credibility.pdf

e.g., planning rolling blackouts that preserve basic function while mitigating economic and human costs, choosing locations for power lines that limit downstream risk, or devising a routing and departure time for a given trip based on energy, emissions, traffic, and other factors.

Building and validating these models—and using them to solve the multivariate, nonlinear optimization problems outlined above—will call upon the full breadth of computing research. In addition to the top-level work in AI, planning, and decision support, developments will also be required throughout the hardware and software stack: new sensing devices and networking strategies, for instance, to build instrumented infrastructure systems. In the case of the cloud—which is projected to consume 5% of US power by 2024[3]—optimizing supply to meet demand will depend on a complete rethinking of resource abstractions (location, intermittence, etc.). The complexity of the associated cost functions is another challenge, and not a purely technical one: what is critical is not total power, for instance, but rather when and where power is consumed, how that maps to supply, and what the costs are—not just the generation costs, but a full accounting of the hidden and downstream costs, including those to individuals, as well as to society, the environment, and the economy. Advances in battery materials and technology will be critical for the reduction of both eWaste and the dependence on the limited sources of rare-earth materials for electronic devices, as well as for scaling up the necessary capacity to fully incorporate renewable energy sources and electric vehicles onto the power grid. Finally, there is a legacy of vulnerable communities bearing the brunt of environmental pollution, for instance, including poor air and water quality—inequities that are intricately interwoven into modern energy systems.

**Environmental Justice**

As the Federal Interagency Working Group on Environmental Justice has noted[4], the legacy of inequity described above includes zoning decisions that have put environmental pollution—incinerators, power plants, sewage facilities, landfills and other waste-disposal mechanisms—in close proximity to low income neighborhoods, and often to communities of color. Simultaneously, such communities have often been deprived of greening measures (e.g., planting trees and other greenery) that can improve air quality, heat management, and quality of life. The combined effects of decades of such practices have further weakened the public health of already vulnerable communities, causing increased instances of asthma, lead-poisoning, etc. Air pollution has also been correlated with decline in mental well-being and even increased suicide rates. As a result, many of these communities experience much worse environmental conditions than wealthy communities, and therefore are already much more vulnerable to significant degradations in public health and wellbeing due to climate change. Similarly, global warming and sea-level rise will differentially impact vulnerable populations worldwide. Moreover, infrastructure improvements that can address sustainability and resilience to climate change have been limited by this inequitable legacy.

---

[3]L. Belkhir. *Assessing ICT global emissions footprint: Trends to 2040 & recommendations.* Journal of Cleaner Production, 2018. https://www.sciencedirect.com/science/article/abs/pii/S095965261733233X#!
[4]*Federal Interagency Working Group on Environmental Justice (EJ IWG)*. United States Environmental Protection Agency (EPA), 2021.
https://www.epa.gov/environmentaljustice/federal-interagency-working-group-environmental-justice-ej-iwg

Addressing environmental justice will require innovations across computing and at the sociotechnical boundary. At the top level, any and all proposed technological advances to address climate change must simultaneously work to address and ultimately overturn this unequal environmental legacy: i.e., they must actively promote *Environmental Justice*. For example, any computing-enabled risk forecasting system, such as an AI-enabled warning system for extreme events (e.g., storms, flooding, wildfire) must prioritize warnings to communities, with the urgency level of warnings to a community based on both its vulnerability and the probability that it is in the "cone of uncertainty" of the extreme event. Such approaches will require the development of rich data sets and models that properly capture and reveal inequities, hidden biases, and economic factors, as well as advances in AI and robotics (expanded on in other sections), for modeling and decision support tools that leverage this data and manage cascading risks (e.g., to manage countermeasures). Moreover, innovations will be needed that bridge Ethics of Computing with Environmental Ethics and Justice.

**Transportation**

Advanced computing technologies bring tremendous promise to advancing and optimizing transportation systems to mitigate impacts of carbon emissions. Opportunities for reductions of carbon emissions span a wide range, from more-efficient personal use vehicles to more-efficient public or shared transportation, via instrumentation of roads, highways, and railways. The design, prototyping, and transition of advanced technologies has the potential to create opportunities for whole-system optimizations. Increased autonomy in our vehicles, complemented by embedding of sensors in transportation networks, could enable real-time route planning and enhanced efficiency. This will require advances in real-time planning, communication systems, and energy-efficient hardware systems. Broad adoption of electric vehicles and associated energy infrastructure would reduce our reliance on traditional fuels. Advanced decision support systems could allow for optimized planning of resources and collaborations with behavioural economists can provide decision makers insights into the most effective financial and behavioral incentives towards shared and public transportation infrastructure and adoption of more efficient technology such as electric vehicles.

Care must also be taken to avoid unfair distribution of these potential benefits. These must include ensuring that communities have consistent access to public infrastructure and that route planning and pricing algorithms do not bias or disadvantage, for example, route recommendations. Advances in privacy preserving computations are needed to ensure that decision support systems based on analysis of aggregated data properly anonymize the data and allow for opt-out mechanisms, while maintaining whole-system efficiency. New methods for secure data sharing can here enable exciting opportunities for public-private partnerships between component agencies of DoT and rideshare companies (as an example) to enable testing and development of new algorithms (e.g., for route planning), decision support systems, and research into socio-technical considerations. Similarly, as with any computing technology, methodologies for effectively recycling, reusing, and updating computing components both in vehicles and in supporting infrastructure (such as highways, railways, and other networks) must be considered in the context of minimizing electronic waste and use of materials.

**Infrastructure**

Increased impacts of climate change and acceleration of frequency of extreme events stress regional power grids, transportation and communications networks, the manufacturing and financial services sectors and other aspects of the nation's critical infrastructure.[5] Examples such as the failure of the Texas power grid in February of 2021 demonstrate challenges impacting our current infrastructure systems. "Black swan events" that exceed infrastructure design specifications are increasingly common. Furthermore, resiliency often requires redundancy and thus potentially reduces efficiency. Security of critical infrastructure is affected both by legacy systems that do not have protections built in and adoption of new technologies without proper protections.

Advances in computing research, along with interdisciplinary collaborations, have the potential to create robust, resilient next-generation infrastructure for the future. Specifically, decision-support systems and advanced algorithms could be developed to predict, identify, and mitigate cascading failures. In addition to leveraging decision-support systems to identify potential tipping points or security flaws in critical infrastructure, decision-support and visualization systems can be leveraged to optimize design of new infrastructure. There is a significant need for balance between protecting and increasing resilience of existing systems while ensuring fair and broad access to new systems as they are deployed. Being able to monitor and optimize heterogeneous infrastructure will be a significant challenge, requiring composable software modules and potentially new programming languages. To be most effective, critical infrastructure will need to communicate and couple to advanced transportation network technologies—again creating opportunities for optimization while also presenting new vulnerabilities. Optimization and planning algorithms can also play a role in the effective reclamation and reuse of materials from decommissioned structures, making the construction industry more sustainable. AI-enabled materials science will also play a role, allowing for discovery of green building materials. Cheap, low-energy sensors could allow for early detection of system failures in power grids; assimilating those data into advanced forecasting and monitoring systems could allow for adaptable and accurate load predictions. Partnerships with scholars in policy and energy management will be needed to design critical infrastructure that is optimized more globally; currently, different methodologies for managing grids across states contribute to failures that are difficult to mitigate. Increased security and resiliency often comes at the cost of efficiency thus requiring development of energy monitoring software systems and energy efficient components.

---

[5] *Critical Infrastructure Sectors*. Cybersecurity & Infrastructure Security Agency, 2020. https://www.cisa.gov/critical-infrastructure-sectors

## Agriculture

Agriculture systems are sensitive to baseline shifts (e.g., decade-scale increases in average annual temperature, increased environmental variability, etc.) as well as to extreme events like droughts, floods, heat waves, and wildfires, which are projected to increase in both intensity and frequency as the climate changes. Anticipating the effects of these conditions and adapting our practices and infrastructure to manage them, will require systems-level models and algorithms that accurately represent and reason about the complex, coupled earth/human system that is modern agriculture. As in many other impact areas mentioned in this paper, these models must be adaptive, data-driven, and spatiotemporal in nature in order to be useful in optimization, planning, and design—e.g., smart crop rotation strategies that are optimized to changing conditions, or spatial distribution of a crop along a wet-dry gradient, making production robust in the face of precipitation variations. They must support risk assessment and incorporate a holistic accounting of economics (e.g., hidden costs of fertilizer runoff, workforce availability). To support multi-factor optimization algorithms in this complex space will require careful consideration of all of the factors that are involved, from the obvious (temperature tolerance) to the subtle (grower trust and acceptance). Rising temperatures allowed the bark beetle to cross the Continental Divide and decimate forests in the Rockies, for instance, and agriculture has roles in both greenhouse gas emission and carbon capture. These models must be designed to facilitate sensitivity analyses for resilience testing and scenario exploration. Moreover, they will need to operate effectively in the context of the smart infrastructure systems (power, water, transportation) that will come online in the near future.

Adaptive multi-factor optimization for rapid re-design and re-planning of agriculture systems in the face of climate change will call upon every subfield of computing research, including software, hardware, and infrastructure to monitor all of the elements of the system (air, water, energy, fire, floods, disease, transportation, supply chains, etc.). Robotics can play important roles here, in both monitoring and execution. We will need new algorithms for processing and using data: quality assurance, cleaning, and fusion, as well as new strategies for assimilating those data into real-time monitoring (nowcasting) and predictive (forecasting) models—as well as automated re-fitting and re-testing strategies for those models. New computing infrastructure is needed that efficiently implement these strategies via in situ sensing and data-driven actuation, automation, and control of farm operations, that is linked to remote, high-performance computing systems when available. AI is critical for adaptive agriculture, for tasks ranging from discovery (e.g., of new crop variants that accommodate climate change-induced state changes) to exploration of scenario spaces for the purposes of both risk assessment and design optimization. AI can also be useful for integrating economics and prediction of unforeseen consequences: e.g., anticipating the negative downstream impacts of weed-control chemicals (pollinator death, evolutionary pressure towards resistant strains) and weighing them against near-term advantages. Finally, we will need analysis, visualization, and decision-support tools to help farmers and policymakers understand and trust these complicated, interwoven, changing problems and solutions.

**Environmental monitoring and forecasting**

Environmental monitoring systems are critical to evaluating the success of any proposed climate mitigation or adaptation measures: both in assessing baselines and tracking improvements. In addition to many of the applications previously discussed, the development of smart environmental monitoring systems will enable other crucial measures in the fight against climate change: for example, in environmental conservation and ecosystem management, pollution source tracking and attribution, and monitoring and characterizing climate change's societal impacts, for example on air quality. Work on these problems, as in the other impact areas described above, will call upon many areas of computing research, including infrastructure, networking technologies, and algorithms for data acquisition, sharing, management, and fusion. Novel AI and data assimilation strategies are needed for real-time model update and adaptation. Autonomous systems and robotic platforms, e.g., UAVs, are also critical for environmental monitoring; examples include measuring greenhouse gasses and adaptively monitoring individual wind turbines. Not only will these endeavors involve developing software, hardware, and computing infrastructure to support environmental sensing, but they will also involve managing the risks of an array of possible system failure modes (e.g., detecting and diagnosing broken sensors). Further, the development of decision support systems with highly accessible user interfaces will significantly reduce the barriers to entry for the human stakeholders—including environmental scientists, members of industry (e.g. transportation, agriculture, and energy), and the general public.

Hand-in-hand with the need for monitoring is the challenge of forecasting. In recent years, many communities have been hit hard by extreme climate and weather events (including extreme storms, flooding, wildfire and resulting debris flows, and drought), which are by definition hard to predict in a changing world. Given the massive amounts of publicly available data, primarily from federally funded sources, computing—especially AI and related technologies—is necessary to unlock insights that will help communities mitigate and adapt to climate change, and prepare for extreme weather events. Societal resilience to climate and environmental risks will require cross-cutting research that combines learning and data with reasoning, decision making, algorithms, and the sociotechnical interface, in order to make better predictions about how essential systems (e.g., supply chains and electric grids) will react to extreme events. For example, the development of risk-sensitive planning and optimization algorithms will be critical for managing response (e.g., hydropower systems in the context of increased rainfall variance, rolling blackouts to protect the grid, pumping stations for NYC subway). In addition to AI, this research will harness and advance the domains of uncertainty quantification; system-level, risk-sensitive modeling, planning, and optimization strategies.

**Conclusion**

Climate change is a quintessential "wicked problem[6]," a complex one with many, often conflicting, interdependencies. The computing research community has the potential to help solve this problem, transforming our climate-change trajectory by enabling resilience, adaptation, and mitigation. There are a number of cross-cutting themes in this work, beginning with understanding, predicting, and controlling complex systems. Many of the impact areas mentioned in this paper, for instance, call for systems that combine real-time data from millions of sensors with historical data, AI models, and "digital twin" simulations to understand system behavior and the impact of various interventions—and then couple to various forms of actuators to effect such interventions. It is important to note that these computational solutions will themselves create additional energy and infrastructure needs; this must be factored into the equation as well, via a holistic analysis of trade-offs and the need for resource- and energy-efficient hardware, software, and algorithms.

---
[6] H. Rittel and M. Webber. *Dilemmas in a General Theory of Planning.* Policy Sciences, *1973.* https://www.jstor.org/stable/4531523

These are grand challenges for computing research. Solutions to these problems will require advances in all areas of computing, carried out by interdisciplinary teams that bring together computing researchers with colleagues from the social, behavioral, and economic sciences—as well as the physical sciences and engineering—and coordination between sectors. For example, development of AI-based techniques that help the general public visualize and understand possible impacts of climate change will require that computer scientists be complemented by psychologists, communication researchers, climate scientists, and economists, with the results disseminated throughout the ecosystem (including academia, government, industry, and the public). Personal smart-phone applications that gather feedback and provide real-time warnings and interventions in order to improve human safety and public health will require participation of computer scientists, electrical engineers, psychologists, public health researchers, public policy researchers, and historians, working in collaboration with infrastructure providers. This type of deep, coordinated interdisciplinary approach will not only facilitate transition of novel research (via stakeholder engagement, for example). More importantly, it will help ensure that innovations are equitable and include safety, robustness, accountability, and fairness from design to deployment.

The material is based upon work supported by the National Science Foundation under Grant No. 1734706. Any opinions, findings, and conclusions or recommendations expressed in this material are those of the authors and do not necessarily reflect the views of the National Science Foundation.